\documentclass[11pt,a4paper]{article}
\usepackage{jcappub}
\usepackage{amssymb}
\usepackage{graphicx}
\usepackage{amsmath}
\usepackage{amsfonts}
\usepackage{hyperref}
\usepackage{subfig}

\title{GRMHD formulation of highly super-Chandrasekhar magnetized white dwarfs: stable configurations of non-spherical white dwarfs}

\author[a]{Upasana Das}
\author[a]{and Banibrata Mukhopadhyay \footnote{Corresponding Author.}}

\affiliation[a]{Department of Physics, Indian Institute of Science, 
Bangalore 560012, India}

\emailAdd{upasana@physics.iisc.ernet.in}
\emailAdd{bm@physics.iisc.ernet.in}

\abstract{The topic of magnetized super-Chandrasekhar white dwarfs is in the limelight, particularly in 
the last few years, since our proposal of their existence. By full-scale general relativistic
magnetohydrodynamic (GRMHD) numerical analysis, we confirm in this work the existence of 
stable, highly magnetized, significantly super-Chandrasekhar white dwarfs with mass more
than $3$ solar mass. While a poloidal field geometry renders the white dwarfs oblate, 
a toroidal field makes them prolate retaining an overall quasi-spherical shape, as speculated
in our earlier work. These white dwarfs are expected to serve as the progenitors of 
over-luminous type Ia supernovae.}


\arxivnumber{1411.5367}

\begin{document}
\maketitle

\section{Introduction} 

Mukhopadhyay and his collaborators initiated exploring the possible existence of highly magnetized, 
significantly super-Chandrasekhar white dwarfs and their 
new mass-limit, much larger than the Chandrasekhar limit 
\cite{kundu,prd12,ijmpd12,prl13,grf13,apjl13,mpla14,jcap14}. 
These white dwarfs appear to be potential candidates for the progenitors of peculiar, overluminous, type Ia supernovae,
e.g. SN~2003fg, SN~2006gz, SN~2007if, SN~2009dc \cite{nature,hicken,yam,scalzo,silverman,taub},
discovered within the past decade, which seem to invoke the explosion of super-Chandrasekhar 
white dwarfs having mass $2.1-2.8M_\odot$. Ostriker \& Hartwick \cite{ost} attempted to model magnetized, rotating 
super-Chandrasekhar white dwarfs in the Newtonian framework.
Adam \cite{adam}, while obtained significantly super-Chandrasekhar Newtonian 
white dwarfs, did not  impose realistic constraints on the fields.
It is particularly
after the initiation by the Mukhopadhyay-group, the topic of 
super-Chandrasekhar white dwarfs has come into the limelight. 

Although some authors 
raised doubts against the existence of these super-Chandrasekhar white dwarfs, 
they were addressed and answered 
in detail in subsequent works (see \cite{mpla14,jcap14} and references therein;
also see \cite{vishal}). In fact,
some other authors, independently, supported the proposal of super-Chandrasekhar white dwarfs
by Das \& Mukhopadhyay \cite{prd12,ijmpd12,prl13,grf13,mpla14,jcap14}, 
based on their respective computations \cite{cheon,herrera1,herrera2}.
However, it is indeed true that all our previous attempts of obtaining highly magnetized 
super-Chandrasekhar white dwarfs
were explored assuming the white dwarfs to be spherical beforehand. While this may be possible 
for a certain magnetic field configuration (as we will
show below), generally, highly magnetized white dwarfs are 
expected to be spheroidal due to magnetic tension. Nevertheless, it is usual in science to begin 
the exploration of any new idea  
with a simpler (sometimes even a toy) model. If the results obtained in the simpler and
idealistic model turn out to be promising, e.g in explaining experimental/observational data, only then 
is it justified to proceed for a more rigorous treatment, in order to develop a concrete scientific theory. 
Hence, being no exception, in our venture to find a fundamental basis behind the formation of 
super-Chandrasekhar white dwarfs,
we proceeded from a simplistic to a more rigorous, self-consistent model
in the following sequence: (1) spherically symmetric Newtonian model with
constant (central) magnetic field, (2) spherically symmetric general
relativistic model with varying magnetic field, (3) a model with
self-consistent departure from spherical symmetry by general relativistic
magnetohydrodynamic (GRMHD) formulation, as we develop in the present work.

In this work, we present the results of GRMHD numerical modeling of static, magnetized white 
dwarfs carried out by using the {\it XNS} code, which is particularly suitable
for solving highly deformed stars due to a strong magnetic field \cite{xns1,pili}. 
However, the {\it XNS} code has so far been used only to compute equilibrium configurations 
of strongly magnetized neutron stars. We appropriately modify this code in order to obtain 
equilibrium configurations of strongly magnetized white dwarfs, for the first time in the literature 
to the best of our knowledge. Moreover, this is the first work exploring full-scale, highly magnetized, 
white dwarfs in a self-consistent general relativistic framework.

This work is organized as follows. In section \ref{setup}, we briefly 
describe the numerical set-up for our formulation. In section \ref{results}, we 
discuss the results of white dwarfs having different magnetic field configurations. 
Finally, we conclude in section \ref{conc}.

\section{Numerical approach and set-up}
\label{setup}

We refer the readers to \cite{xns1} and \cite{pili} 
for a complete description of the GRMHD equations, which include the equations characterizing the geometry 
of the magnetic field and the underlying current distribution, as well as the numerical technique 
employed by the {\it XNS} code to solve them. Here we briefly describe mainly those aspects of our numerical set-up which 
differ from the neutron star set-up and is characteristic to our white dwarf solutions.

The presence of a strong magnetic field in a compact object generates an anisotropy in the 
magnetic pressure which in turn causes it to be deformed \cite{monica,bocquet}. 
Of course, the degree of anisotropy depends on the strength and geometry of the magnetic field.
We construct axisymmetric white dwarfs in spherical polar coordinates $(r,\theta,\phi)$, 
self-consistently accounting for the deviation from spherical symmetry due to a strong 
magnetic field. The computational grid along the radial coordinate $r$ extends from an 
inner boundary $R_{\rm min}=0$ to an outer boundary $R_{\rm max}$, while the polar angular 
coordinate $\theta$ is bounded as $0 \leq \theta \leq \pi$. A uniform grid is used along 
both the co-ordinates. Note that $R_{\rm max}$ has to be set such that it is always larger than 
the radius of the white dwarf, which in turn depends on its central density and central magnetic field strength. 
In this work, $R_{\rm max}$ typically ranges from $1000-10,000$ km.
We define the surface of the white dwarf such that $\rho_{\rm surf}=10^{-7}-10^{-8}\rho_c$, 
where $\rho_{\rm surf}$ and $\rho_c$ are the surface and central densities of the white dwarf respectively. 
Our radial boundary conditions are same as those described by \cite{pili}. The number of grid points 
along $r$ and $\theta$ are typically set to be $N_r=500$ and $N_\theta=100$ respectively.
Higher resolution runs (say, with $N_r=1000$ and $N_\theta=500$) only require more computational 
time without causing any significant change in the results.

In this work, we are interested in obtaining equilibrium solutions of such high density, magnetized, 
relativistic white dwarfs, which can be 
described by a polytropic equation of state (EoS) $P=K\rho^\Gamma$, with an adiabatic index $\Gamma \approx 4/3$ 
and the constant $K=(1/8)(3/\pi)^{1/3} hc/(\mu_e m_H)^{4/3}$, where $P$ is the pressure, $\rho$ the density, $h$ 
Planck's constant, $c$ the speed 
of light, $\mu_e$ the mean molecular weight per electron ($\mu_e=2$ here) and $m_H$ the mass of hydrogen atom \cite{chandra35}.
Hence, we neglect the possible effect of Landau quantization on the above EoS which could arise due to a strong magnetic field 
$B>B_c$, where $B_c=4.414\times10^{13}$ G, is a critical magnetic field.
We recall that the maximum number of Landau levels occupied by electrons in the presence of a magnetic field 
is given by 
\begin{equation}
\nu_m = \frac{\left(\frac{E_{Fmax}}{m_ec^2}\right)^{2} - 1}{2B/B_c},
\end{equation}
where $m_e$ is the mass of the electron and 
$E_{Fmax}$ the maximum Fermi energy of the system, 
which is directly related to $\rho_c$ \cite{prd12}. We find that 
for a range $10^{10} \lesssim \rho_c \lesssim 10^{11}$ gm/$\rm cm^3$ considered here (typical to white dwarfs of our interest), the maximum 
magnetic field strength inside the white dwarf ranges as $10^{13} \lesssim B_{\rm max} \lesssim 10^{15}$ G. 
Consequently, $\nu_m \gtrsim 20$ for this range of $\rho_c$ and $B_{\rm max}$, which does not 
significantly modify the value of $\Gamma$ we choose (e.g. see Fig. 4(a) of \cite{prd12}), hence justifying our assumption.

\section{Results}
\label{results}

In this section, we explore the effect of different magnetic field 
configurations on the structure and properties of white dwarfs. 

\subsection{Purely toroidal, purely poloidal and mixed magnetic field configurations}

\begin{figure*}[h]
\centering
\includegraphics[scale=0.5]{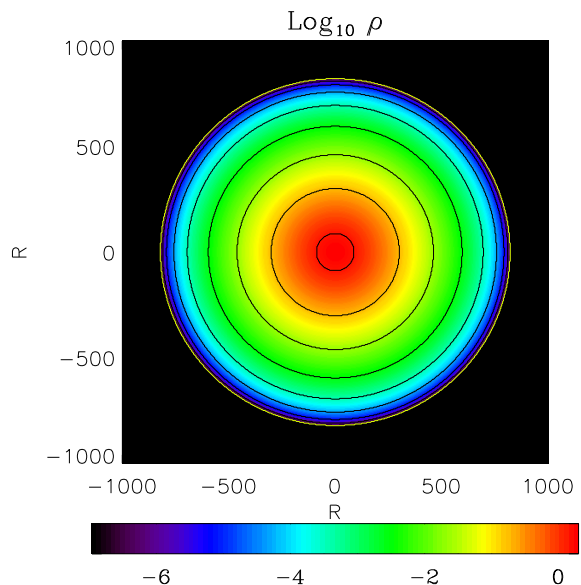}
\caption{Iso-contours of baryonic density of a non-magnetized white dwarf. R is in units of 
$GM_\odot/c^2=1.476$ km and $\rho$ in units of $10^{10}$ gm/cm$^3$. The yellow curve represents the stellar surface.}
\label{nonmag}
\end{figure*}

First, as a fiducial model, we choose a non-magnetized white dwarf with $\rho_c=2\times 10^{10}$ gm/$\rm cm^3$, 
for comparison with the magnetized cases. It has a baryonic mass\footnote{For the definitions 
of all global physical quantities characterizing the solutions, we refer to Appendix B of \cite{pili}} 
$M_0 = 1.416 M_\odot$ and equatorial radius $R_{eq}=1221.94$ km, 
and is perfectly spherical in structure with $R_p/R_{eq}=1$, $R_p$ being the polar radius, as shown in Figure \ref{nonmag}.

\captionsetup[subfigure]{position=top}
\begin{figure*}
  \centering
    \begin{tabular}{ll}
 \subfloat[]{\includegraphics[scale=0.35]{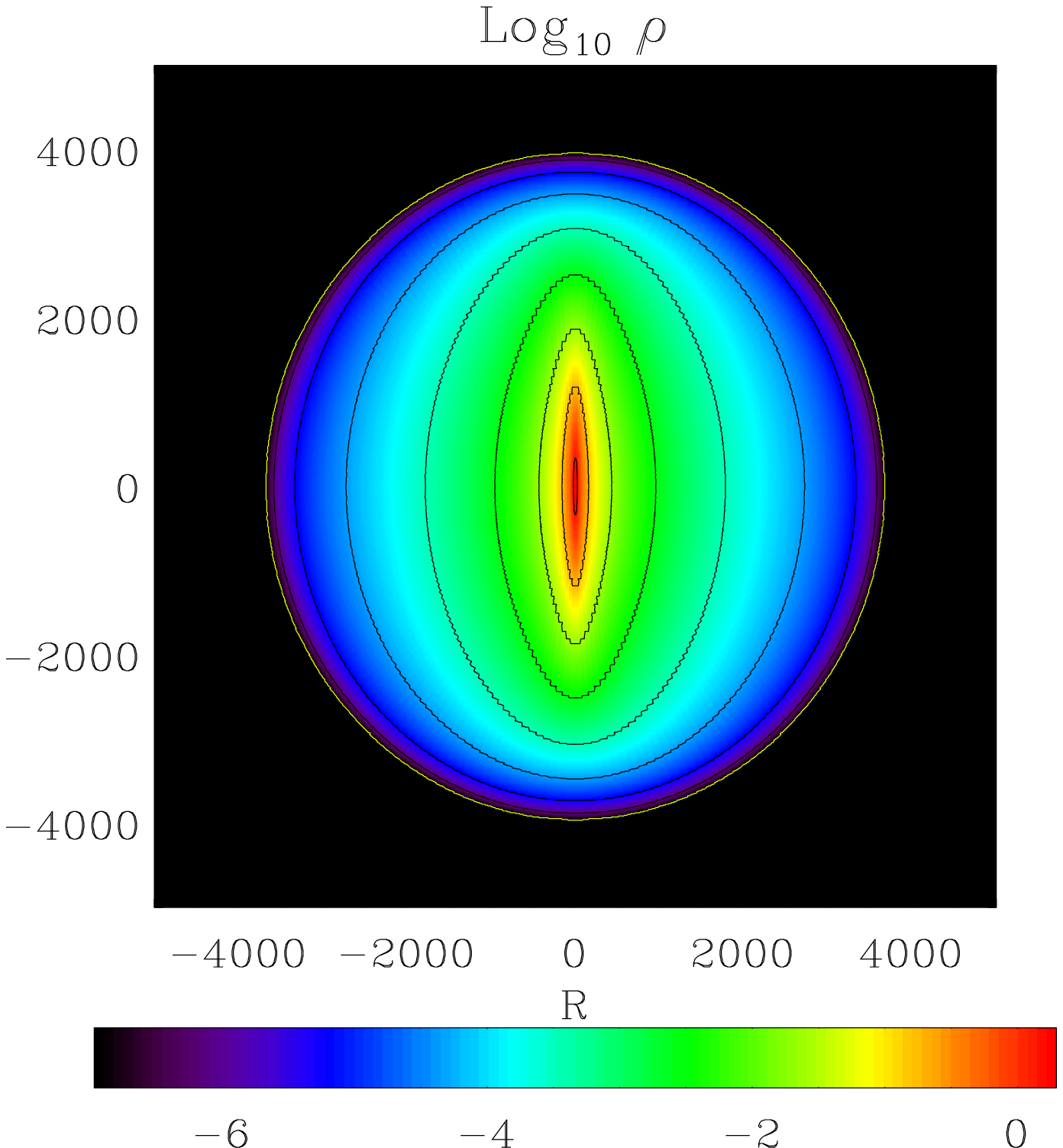}}&
   \subfloat[]{\includegraphics[scale=0.35]{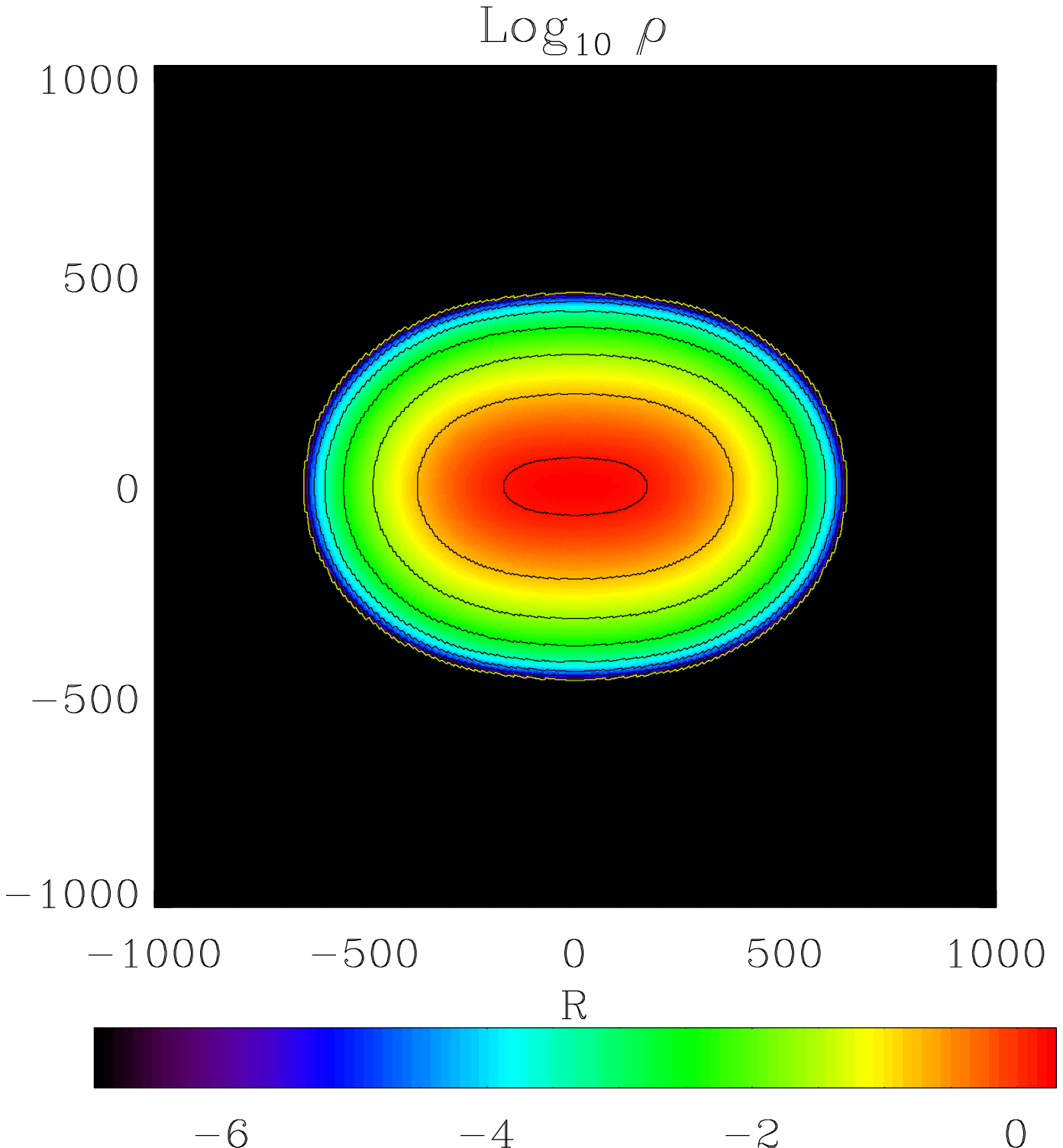}} \\ \\ 
    \subfloat[]{\includegraphics[scale=0.35]{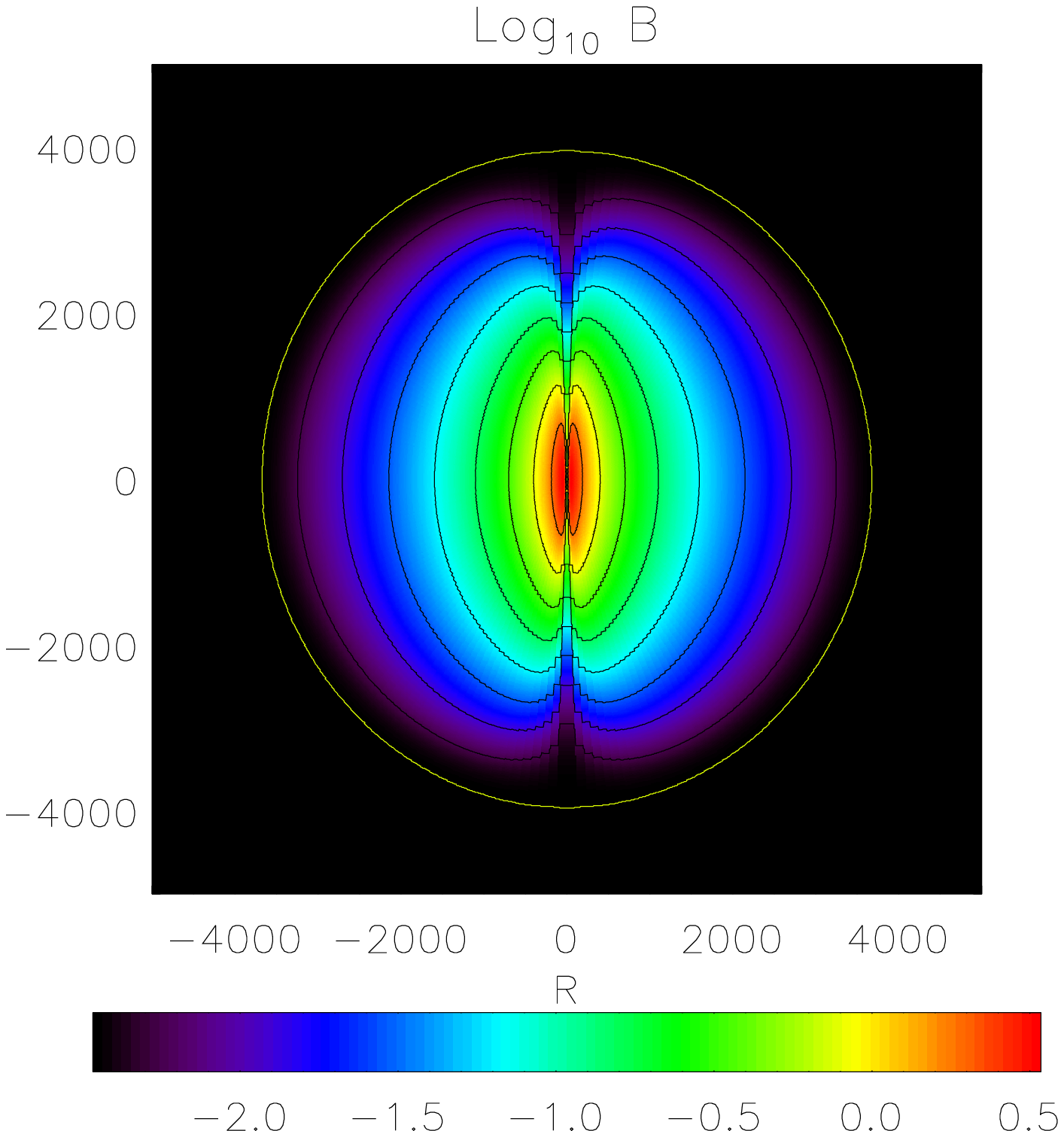}}&
 \subfloat[]{\includegraphics[scale=0.35]{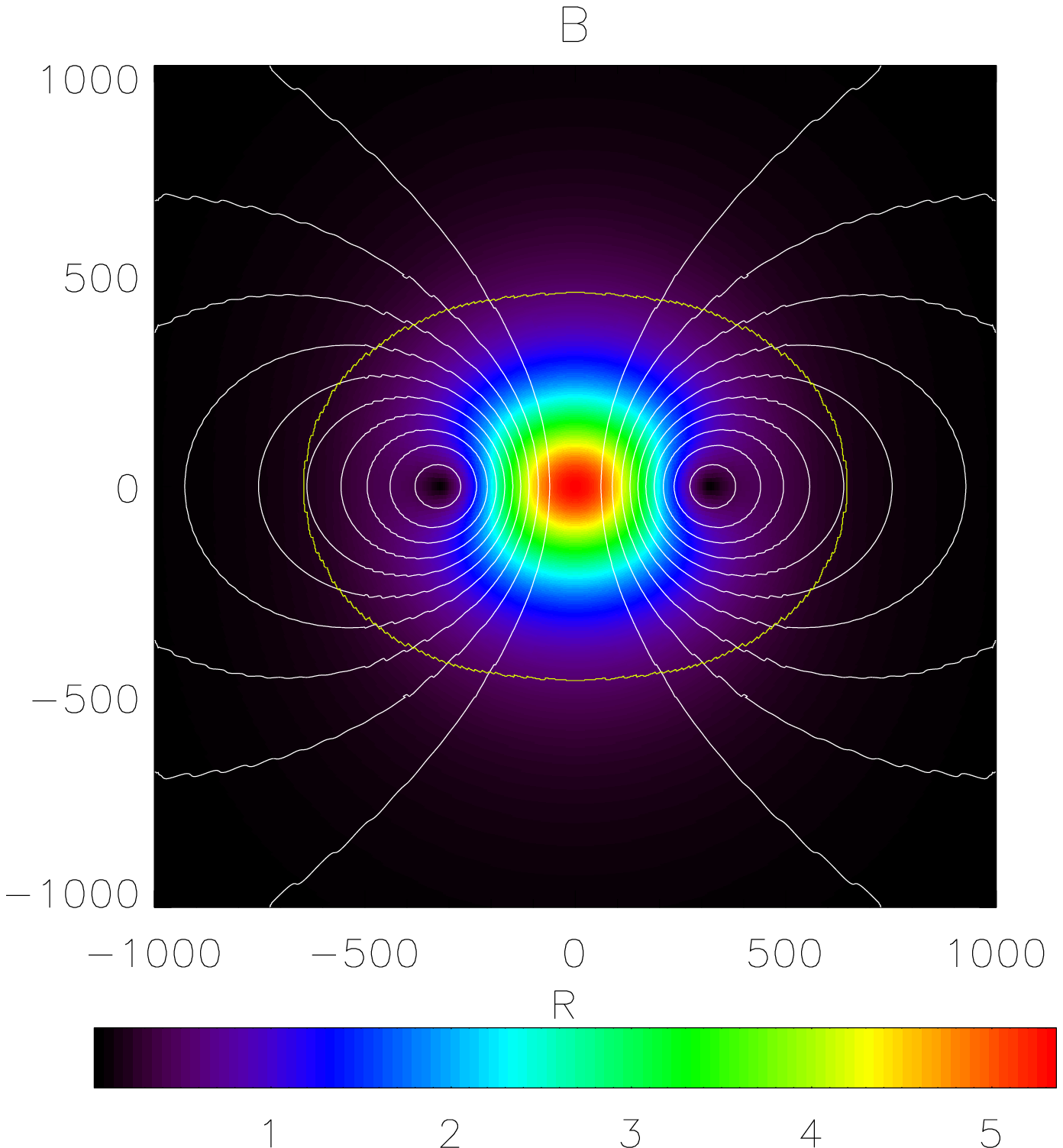}} \\
      \end{tabular}
        \caption{Purely toroidal configuration: iso-contours of (a) baryonic density and (c) magnetic field strength $B=\sqrt{B_\phi B^\phi}$. 
Purely poloidal configuration: iso-contours of (b) baryonic density and (d) magnetic field strength $B=\sqrt{B_r B^r + B_\theta B^\theta}$, 
superimposed with dipolar magnetic field lines (in white). R is in units of $GM_\odot/c^2=1.476$ km, $\rho$ in units of $10^{10}$ gm/cm$^3$
and $B$ in units of $10^{14}$ G. 
The yellow curve in each panel represents the stellar surface.}
\label{poltor}
\end{figure*}

Figures \ref{poltor}(a) and (c) show the distribution of baryonic density and 
magnetic field strength respectively, for a white dwarf with $\rho_c=2\times 10^{10}$ gm/$\rm cm^3$, 
having a purely toroidal magnetic field configuration $\vec{B}= B_\phi \hat{\phi}$. This case corresponds 
to $m=1$ and $K_m=5.37$ in equation (39) of \cite{pili}, where $K_m$ is the toroidal magnetization constant and $m$ is
the toroidal magnetization index. The maximum magnetic field strength attained inside the white dwarf is 
$B_{\rm max}=3.41 \times 10^{14}$ G. Interestingly, this white dwarf is highly super-Chandrasekhar, having $M_0=3.413M_\odot$ 
and $R_{eq}=5463.65$ km. The value of the surface magnetic field does not
affect this result, as long as it is less than $\lesssim 10^{11}$ G, which is the case here, when 
white dwarfs have been observed with surface field as high as $\sim 10^9$  G so far.
More importantly, the ratio of the total magnetic energy to the total gravitational 
binding energy, $E_{mag}/E_{grav}=0.3045$ (which is much $<1$). 
The radii ratio, $R_p/R_{eq}=1.074$ (which is slightly $> 1$), indicating a net prolate deformation in the shape caused due to a 
toroidal field geometry. 
White dwarfs with even smaller $E_{mag}/E_{grav}$ are also found to be highly super-Chandrasekhar 
(see Figs. \ref{rhocf}a and d). This argues for the white dwarfs to be stable (see, e.g. \cite{ost}). 
Figure \ref{poltor}(a) shows that although the iso-density contours close to the center are compressed 
into a highly pronounced prolate structure, the external layers expand, giving rise to an 
overall quasi-spherical shape. Interestingly, this justifies the spherically symmetric assumption 
in our recent work \cite{jcap14}, where we obtained stable 
general relativistic solutions, in the presence of varying magnetic field, of super-Chandrasekhar 
white dwarfs with maximum mass $\sim 3.3M_\odot$. Thus, one can conclude that even spherical, magnetized white dwarfs 
can be realistic and can closely resemble solutions obtained from more accurate considerations, provided their 
magnetic field is appropriately distributed.

Figures \ref{poltor}(b) and (d) show the distribution of baryonic density and 
magnetic field strength superimposed by magnetic field lines respectively, 
for a white dwarf with $\rho_c=2\times 10^{10}$ gm/$\rm cm^3$, 
having a purely poloidal magnetic field configuration $\vec{B}=B_r \hat{r} + B_\theta \hat{\theta}$. This case 
corresponds to $k_{\rm pol}=0.01109$ and $\xi=0$ in equation (32) of \cite{pili}, 
where $k_{\rm pol}$ is the poloidal magnetization constant 
and $\xi$ is the nonlinear poloidal term. Note that $\xi=0$ ensures the presence of linear currents only, 
thus leading to a purely dipolar magnetic field.
The maximum magnetic field strength attained at the center, in this computation, is $B_{\rm max}=5.34\times 10^{14}$ G, 
which also leads to a significantly super-Chandrasekhar white dwarf having $M_0=1.771M_\odot$ 
and $R_{eq}=956.14$ km. The white dwarf is highly deformed with an overall oblate shape and 
$R_p/R_{eq}=0.7065$, which is expected to be stable because of its $E_{mag}/E_{grav}=0.1138$, 
which is very much $<1$ \cite{ost}. In this context we mention that in a recent work, 
a maximum mass of $1.9M_\odot$ has been obtained for a white dwarf having a purely poloidal 
field configuration with interior field $10-100 B_c$, in a Newtonian framework \cite{bera}. Although 
a direct comparison of our work with theirs is not possible, we note that in the presence of a strong 
magnetic field, as both the works choose, general relativistic effects can not be neglected. In fact, we already 
emphasized the importance of general relativistic effects in strongly magnetized white dwarfs in our earlier work \cite{mpla14}, 
which argued that the inclusion of magnetic density could decrease the mass of the white dwarf. Hence, the Newtonian 
results obtained in \cite{bera} appear to be incomplete.
Indeed, the present work shows that general relativistic effects seem to cause a significant decrease in the white dwarf mass
in the purely poloidal case compared to the results of \cite{bera}.

\captionsetup[subfigure]{position=top}
\begin{figure*}[h]
\centering
    \begin{tabular}{ll}
\subfloat[]{\includegraphics[scale=0.4]{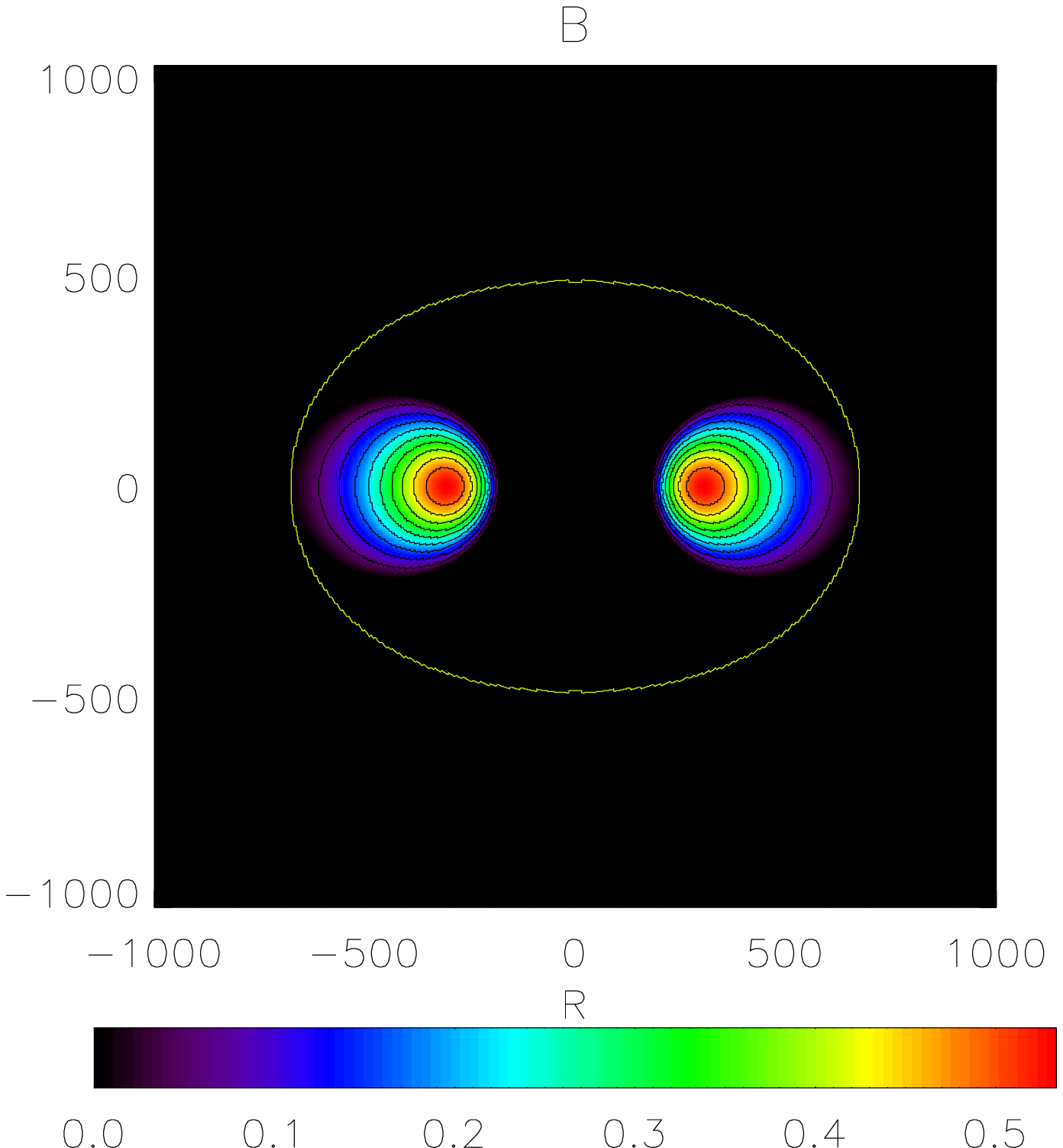}}&
\hspace*{-0.8cm} 
  \subfloat[]{\includegraphics[scale=0.4]{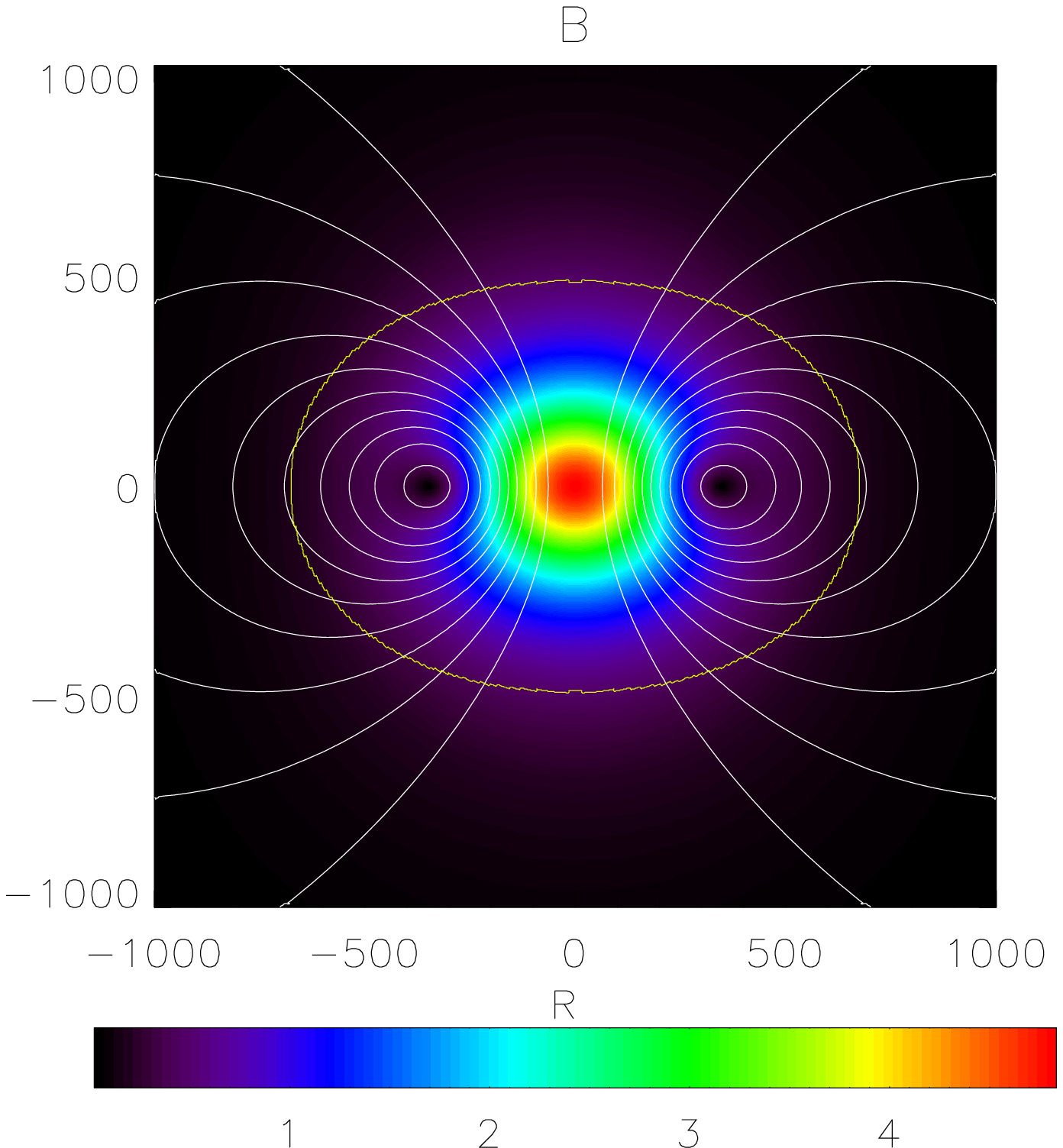}} \\ 
      \end{tabular}
\caption{Twisted torus configuration: magnetic field strength of (a) toroidal component and  (b) poloidal component.
R is in units of $GM_\odot/c^2=1.476$ km and $B$ in units of $10^{14}$ G.
The yellow curve in each panel represents the stellar surface.
}
\label{mixed}
\end{figure*}

Purely toroidal and poloidal field configurations 
are believed to suffer from MHD instabilities (e.g. \cite{tayler}), which 
might result in a rearrangement of the field into a mixed configuration (e.g. \cite{braith}). 
Hence, for completeness, we also construct equilibrium models of 
white dwarfs harboring a mixed magnetic field configuration. 
In the present work, we consider a special case, namely the 
twisted torus configuration. Such a configuration is dominated by the poloidal field, while the toroidal field 
traces only a narrow ring-like region in the equatorial plane close to the stellar surface. 
In Figure \ref{mixed}, we present the results for a white dwarf with 
$\rho_c=2\times 10^{10}$ gm/$\rm cm^3$, having $k_{\rm pol}=0.01$, $\xi=0$ in equation (32) and $a=0.00417$, $\zeta=0$ 
in equation (33) of \cite{pili}, which characterize a twisted torus configuration, such that 
$a$ is the twisted torus magnetization constant and $\zeta$ is the
twisted torus magnetization index. This again results in a significantly super-Chandrasekhar white 
dwarf having $M_0=1.754M_\odot$ and $R_{eq}=1000.44$ km. Figures \ref{mixed}(a) and (b) compare the distribution 
of the toroidal and poloidal components of the magnetic field respectively. 
The poloidal component has a structure very similar to the purely poloidal 
case (Fig. \ref{poltor}d), covering almost the entire interior and 
attaining a maximum at the center $B_{\rm max}=4.82 \times 10^{14}$ G. 
On the contrary, the toroidal component is an order of magnitude smaller than the poloidal component, 
restricted only to a torus in the outer low-density layers of the white dwarf. 
The toroidal field attains a maximum value of $5.34\times 10^{13}$ G, exactly in the region where the poloidal field vanishes. The energy in the toroidal component is only $\sim 2.4\%$ of the total magnetic energy. 
However, note that the toroidal component in this case appears to occupy a larger area 
compared to that in neutron stars (see, e.g., Fig. 12 of \cite{pili}).
Since this mixed configuration is clearly
dominated by the poloidal component of the field, the baryonic density distribution 
looks very similar to that in the purely poloidal case (Fig. \ref{poltor}b) and the white dwarf assumes a highly 
oblate shape with $R_p/R_{eq} = 0.719$. This white dwarf is also expected to be stable having $E_{mag}/E_{grav}=0.1126<1$. In a toroidal dominated 
mixed field configuration \cite{ciolfi13} or in a configuration with similar contributions from both poloidal and 
toroidal components, more massive super-Chandrasekhar white dwarfs could be possible.

Note that the nature of deformation induced in white dwarfs, due to purely 
toroidal, purely poloidal and mixed magnetic field configurations, closely resemble that observed in 
magnetized neutron stars \cite{pili}.

\begin{figure*}[h]
\centering
\includegraphics[scale=0.43]{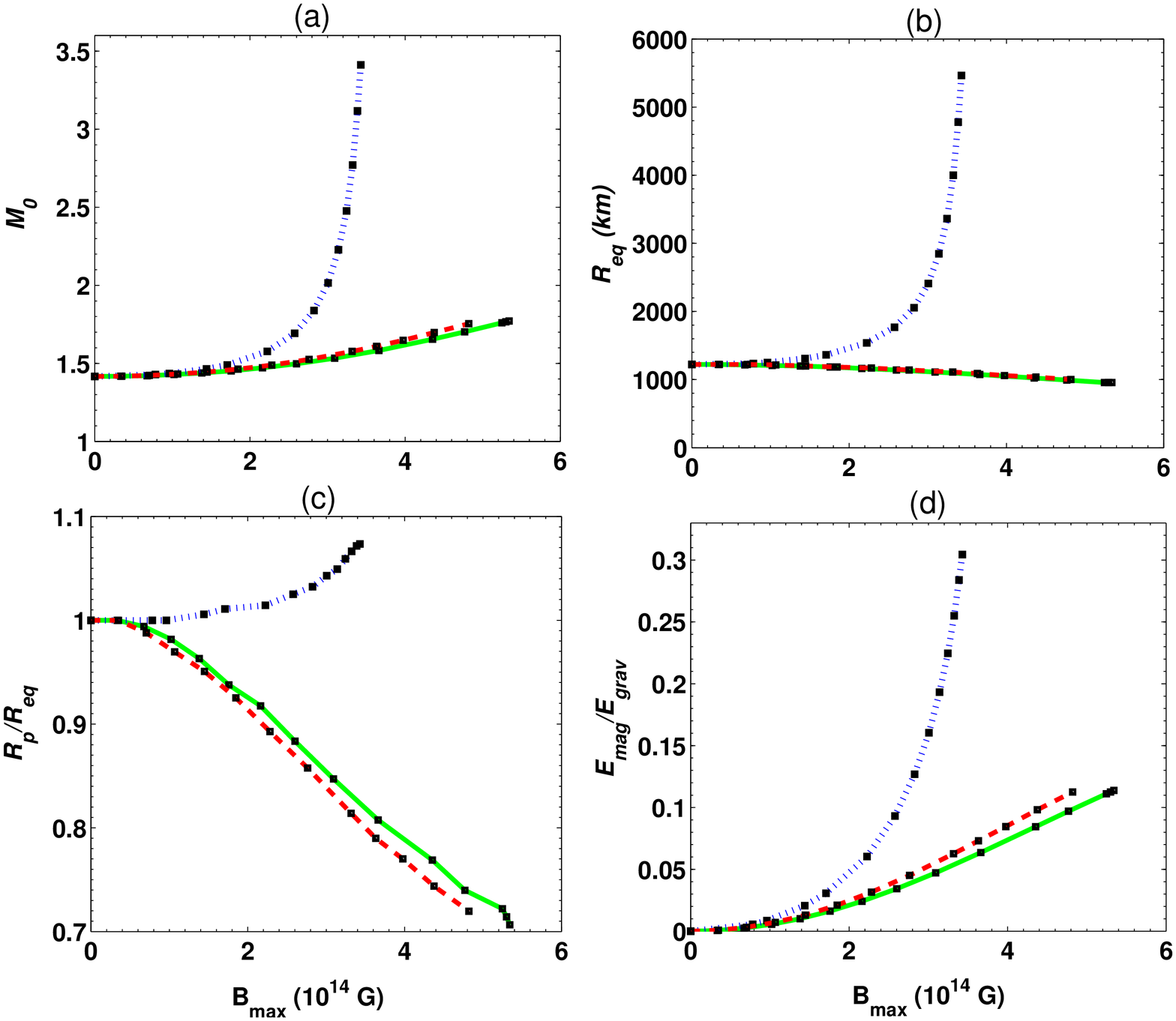}
\caption{Equilibrium sequences of magnetized white dwarfs with fixed $\rho_c=2\times 10^{10}$ gm/$\rm cm^3$. (a) $M_0$, (b) $R_{eq}$, (c) $R_p/R_{eq}$ 
and (d) $E_{mag}/E_{grav}$, as functions of $B_{\rm max}$. The solid (green), dotted (blue) and dashed (red) 
curves represent, respectively, 
white dwarfs having purely poloidal, purely toroidal, and twisted torus field configurations. $M_0$ is in units of $M_\odot$. 
The filled boxes represent individual white dwarfs.
  }
\label{rhocf}
\end{figure*}

\subsection{Equilibrium sequences at fixed central density}

We also construct equilibrium sequences of magnetized white dwarfs pertaining to different field 
geometries, for a fixed central density $\rho_c=2\times 10^{10}$ gm/$\rm cm^3$ but corresponding to 
different $B_{\rm max}$. Figure \ref{rhocf} shows the variations of different physical quantities as functions of $B_{\rm max}$.

Figure \ref{rhocf}(a) shows an increase in the white dwarf mass with an increase in magnetic field for all the 
three field configurations discussed above, eventually leading 
to highly super-Chandrasekhar white dwarfs. The purely toroidal case shows a steeper increase in the mass compared 
to the other field configurations. 
Note that the poloidal field domination in the twisted torus configuration causes its curve to closely 
trace that of the purely poloidal case.

Figure \ref{rhocf}(b) shows an increase in $R_{eq}$ with the increase in magnetic field for the purely 
toroidal case. However, the situation reverses if a poloidal field component is present. $R_{eq}$ decreases with the increase in magnetic field for the purely 
poloidal and twisted torus cases.

Figure \ref{rhocf}(c) shows that with the increase in magnetic field, the white dwarfs become more deformed in shape, 
characterized by $R_p/R_{eq}$ deviating increasingly from unity. The purely 
toroidal cases exhibit a prolate deformation and hence $R_p/R_{eq}>1$. The purely poloidal 
and twisted torus cases exhibit an oblate deformation with $R_p/R_{eq}<1$. Note, however, that the 
departure from spherical symmetry is more pronounced in the purely poloidal (and hence twisted torus) cases compared 
to that in the purely toroidal cases.

Finally, Figure \ref{rhocf}(d) shows that with the increase in magnetic field, the magnetic energy in the white dwarf 
increases, as expected, for all the three field configurations. Nevertheless, the magnetic energy 
remains (significantly) sub-dominant compared to the gravitational binding energy for all the cases, 
since $E_{mag}/E_{grav}<1$ always.
This, very importantly, argues for the possible existence of stable, 
highly magnetized super-Chandrasekhar white dwarfs.

\section{Summary and Conclusions}
\label{conc}

By carrying out extensive, self-consistent, GRMHD numerical analysis of magnetized white dwarfs, for the first time 
in the literature, we have 
reestablished the existence of highly super-Chandrasekhar, stable white dwarfs. 
Our main message since the last couple of years has been that the enormous power of magnetic field ---
irrespective of its nature of origin: quantum, classical and/or general relativistic --- is capable of revealing
significantly super-Chandrasekhar white dwarfs, which is further strengthened in this work.
Our first set of works in this venture demonstrated the power of quantum effects, revealing super-Chandrasekhar white 
dwarfs with mass up to $2.58M_\odot$ \cite{prd12,ijmpd12,prl13,grf13}. Thereafter, Das \& Mukhopadhyay \cite{jcap14} showed the 
power of general relativistic effects along with magnetic pressure gradient, revealing mass up to $3.3M_\odot$ --- again
significantly super-Chandrasekhar. Finally, the present GRMHD analysis has confirmed the 
above proposals of the existence of
super-Chandrasekhar white dwarfs.
These white dwarfs can be ideal progenitors of the peculiar, overluminous type Ia supernovae.

A strong magnetic field causes the pressure to become anisotropic, 
which in turn causes the underlying white dwarf to be deformed in shape. 
In order to self-consistently study this effect, we have explored 
various geometrical field configurations, namely, purely toroidal, purely poloidal and twisted torus 
configurations. Interestingly, we have obtained super-Chandrasekhar white dwarfs with mass $1.7-3.4M_\odot$ 
even for relatively low magnetic field strengths, when the deviation from spherical 
symmetry is taken into account --- as already speculated 
based on the simple computation given in the Appendix of \cite{prd12}.
This confirms the importance of general relativistic effects in highly magnetized white dwarfs.

We have furthermore noted that for a fixed $\rho_c$, a higher magnetic field yields a 
larger mass and a higher deformation for all field configurations. A purely poloidal 
field yields oblate equilibrium configurations. The characteristic deformation 
induced by a purely toroidal field is prolate, however, the overall shape remains quasi-spherical, 
justifying the earlier spherically symmetric assumption in computing models of strongly 
magnetized white dwarfs \cite{prd12,prl13,jcap14}.
We have also considered mixed field or twisted-torus configurations, where the poloidal 
component is energetically dominant, yielding oblate white dwarfs similar to the purely poloidal case.
Computing mixed field models with toroidal dominated configurations is somewhat more challenging 
and has only recently been demonstrated by Ciolfi \& Rezzolla \cite{ciolfi13}. It would be interesting to 
investigate the effect of such a field configuration on white dwarfs, as a toroidal domination 
is expected to yield even more massive white dwarfs compared to the twisted torus case.

\acknowledgments

We thank J. P. Ostriker for continuous encouragement and appreciation. We are also truly grateful to 
N. Bucciantini and A.G. Pili for their extremely helpful  inputs 
regarding the execution of the {\it XNS} code.
B.M. acknowledges partial support through research Grant No. ISRO/RES/2/367/10-11. 
U.D. thanks CSIR, India for financial support.

\end{document}